\documentclass[11pt,twoside]{article}
\usepackage{cspm-asp2006}
\usepackage{epsfig,graphicx,natbib,url}  
\usepackage{lscape}
\begin{document}

\markboth{Hudson}{Chromospheric Flares} 
\title{Chromospheric Flares} 
\author{Hugh S. Hudson} 
\affil{Space Sciences Laboratory, University of California, Berkeley} 

\begin{abstract} 
In this topical review I revisit the ``chromospheric flare''.  This
should currently be an outdated concept, because modern data 
seem to rule out
the possiblity of a major flare happening independently in the
chromosphere alone, but the chromosphere still plays a major
observational role in many ways.  It is the source of the bulk of a
flare's radiant energy -- in particular the visible/UV continuum
radiation.  It also provides tracers that guide us to the coronal
source of the energy, even though we do not yet understand the
propagation of the energy from its storage in the corona to its
release in the chromosphere.  The formation of chromospheric
radiations during a flare presents several difficult and interesting
physical problems.
\end{abstract}

\section{Introduction}

Solar flares first revealed themselves as visual perturbations of the
solar atmosphere (``white light flares'') and hence immediately were
construed as a photospheric process.  With the invention of
spectroscopic techniques, though, it became clear that chromospheric
emission lines such as H$\alpha$ revealed flare presence much more
readily.  This led to the concept of the ``chromospheric flare'' and
to a great deal of observational material on H$\alpha$ flares and
eruptions, as reviewed by \citet{hudson-1963QB528.S6.......},
\citet{hudson-1966soat.book.....Z}, or
\citet{hudson-1976sofl.book.....Svestka}, for example.  At some point,
prior to the discovery of coronal flare effects, the
misinterpretations of the H$\alpha$ line profile even led to the
incorrect idea that a flare was a sudden {\it cooling} of the solar
atmosphere.  In any case, a perturbation of the lower solar atmosphere
violent enough to affect the solar luminosity itself (``white light") 
implies a large energy content.

Our view of flares now emphasizes the high temperatures and
non-thermal effects seen in the corona, and we generally believe the
chromospheric effects themselves to be secondary in nature.  This may
be true, but nonetheless the modern observations confirm the fact that
the lower solar atmosphere dominates the radiant energy budget of a
flare via the UV and white-light continua.  Somehow, therefore, the
energy stored in the solar corona rapidly focuses down into regions
visible in chromospheric signatures; this accounts for the high
contrast of flare effects there.  Thus the ``chromospheric flare''
remains essential to our understanding of the overall processes
involved.

The chromosphere nowhere exists as a well-defined layer with a
reproducible height structure.  In this paper I use the term
interchangeably with ``lower solar atmosphere,'' embracing the
phenomena of the visible photosphere through the transition region.
During flares the structure of these ``layers'' and the physical
conditions within them may change drastically.  The changes generally
happen so fast and on such small spatial scales that we cannot observe
them comprehensively.  Understanding the impulsive phase in the
chromosphere may therefore seem like something of a lost cause from
the the point of view of theory, especially in view of our inability
to understand the {\it quiet} chromosphere any better than we do.  The
data repeatedly reveal that we simply have not yet
resolved the spatial or temporal structures involved in the impulsive
phase, and that without knowing the geometry of the physical
structure, we cannot really comprehend its physics.  The TRACE
(\cite{hudson-1999SoPh..187..229Handy}) and RHESSI
(\cite{hudson-2002SoPh..210....3L}) observations have provided more
than one recent breakthrough, however, and it may be that we are
beginning to understand the {\it gradual} phase of a flare at least.

This review is organized around several topics involving the behavior
of the chromosphere during a flare.  These include the process of
``chromospheric evaporation'' (Section~\ref{hudson-sec:evaporation}),
flare energetics (Section~\ref{hudson-sect:ergs}), the mechanisms of
flare continuum emission (Section~\ref{hudson-sec:cont}), and the
inference of flare structure from the morphology of the chromospheric
flare (Section~\ref{hudson-sec:struct} and
Section~\ref{hudson-sec:surge}).  In Section~\ref{hudson-sec:hist} and
Section~\ref{hudson-sec:spectrum} we give an overview of the history
of chromospheric flares and show a cartoon to establish a working
model of a solar flare.  Sections~\ref{hudson-sec:recon}
and~\ref{hudson-sec:model} discuss large-scale magnetic reconnection
and theoretical ideas, and Section~\ref{hudson-sec:gamma} presents a
$\gamma$-ray mystery.

\section{Historical Development}\label{hudson-sec:hist}

\begin{figure}
\centering
\includegraphics[width=0.45\linewidth]{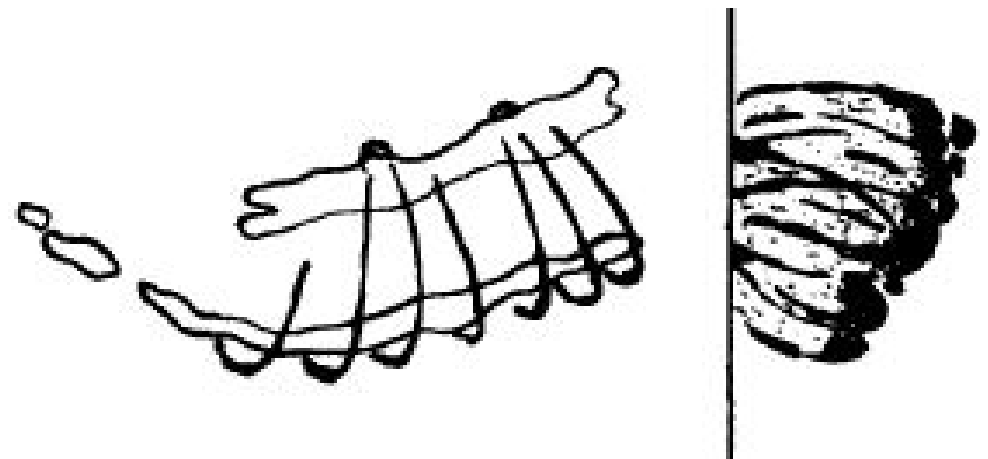}
\includegraphics[width=0.45\linewidth]{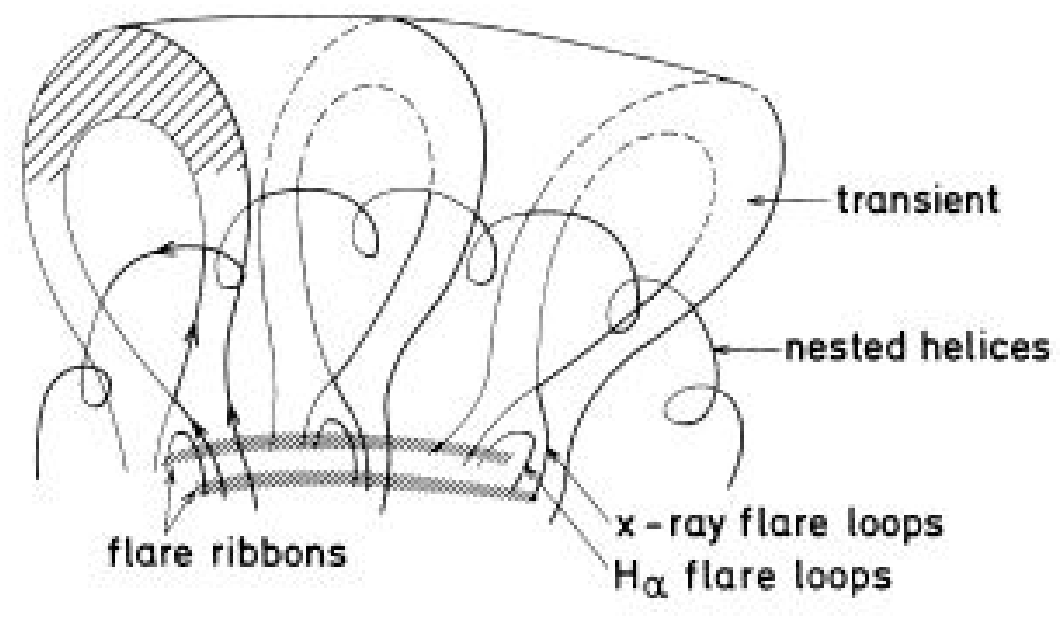}
\caption{Left: one of Bruzek's (1964) sketches, showing a flare with
ribbons on the disk and its equivalent H$\alpha$ ``loop prominence
system'' over the limb.  This key observational pattern led directly
to the formation of our standard flare model (right), in the form
presented by Anzer \& Pneuman (1982).  }
\label{hudson-fig:bruzek}
\end{figure}

Although it was the white-light continuum that initially revealed the
existence of solar flares, the advent of spectroscopy (e.g.,
\cite{hudson-1930ApJ....71...73H}) allowed their regular observation
via the H$\alpha$ line (see \cite{hudson-1966SSRv....5..388S} for a
discussion of the historical development of these observations).  This
strong absorption line actually becomes an emission line during bright
flares, and H$\alpha$ limb observations frequently show prominences
and eruptions.  H$\alpha$ observers came to recognize a particular
flare morphology, the so-called two-ribbon flare.
\citet{hudson-1964ApJ...140..746B} described the patterns followed by
these events, which provided strong evidence that the solar corona had
to play a major role in flare development.
Figure~\ref{hudson-fig:bruzek} reproduces one of Bruzek's sketches,
and then illustrates in a cartoon (due to
\cite{hudson-1982SoPh...79..129A}) how this morphology led to our
standard magnetic-reconnection scenario that 
tries to embrace the X-ray
observations and the coronal mass ejections (CMEs) as well as the
chromospheric ribbon structures.

In this standard picture a solar flare develops in a complicated
manner that involves restructuring of the coronal magnetic field in
such a way as to release energy.  The immediate effects of this energy
release are to produce broad-band ``impulsive phase'' emissions and to
drive chromospheric gas up into coronal magnetic loops, the process we
term ``chromospheric evaporation.''  A part of the field magnetic
structure may actually erupt and open out into the solar wind, in the
sense that the field lines stretch out past the Alfv{\'e }n critical
point of the flow.  This opening may consist of rising loops which
then take the form of a coronal mass ejection (CME), or it may involve
interactions with previously open field (a process often termed
``interchange reconnection'' nowadays; see
\cite{hudson-1977ApJ...216..123H}).  If a CME does accompany the
flare, as it almost invariably does for flares of GOES class~X or
greater, the energy involved in mass motions may be comparable to the
luminous energy (e.g., \cite{hudson-2005JGRA..11011103E}).  Generally
the observations are limited in resolution, both temporal and spatial,
and especially in spectral coverage.  Thus we often resort to a
cartoon that serves to identify how the essential parts of a flare
relate to one another.

Soft X-ray observations show hot loops in the gradual phase of a
flare.  These result from the material ``evaporated'' from the
chromosphere and have anomalously high gas pressure (but still low
plasma $\beta$; however see \cite{hudson-2001SoPh..203...71G}).
Whereas the pressure at the base of the corona normally is of order
0.1~dyne~cm$^{-2}$, a bright flare loop can achieve
10$^{3}$~dyne~cm$^{-2}$.  This over-dense and over-hot coronal loop
gradually cools, and in its final stages the remaining plasma returns
to a more chromospheric state and suddenly becomes visible in
H$\alpha$ (\cite{hudson-1971SoPh...19...86G}).  The loops that have
reached this state then form Bruzek's H$\alpha$ loop prominence system
(Figure~\ref{hudson-fig:bruzek}).

During the ribbon expansion another important phenomenon occurs: hard
X-ray emission appears at the footpoints of the coronal loops that are
in the process of being filled by chromospheric evaporation
\citep{hudson-1981ApJ...246L.155H}.  The hard X-rays show that a
substantial part of flare energy appears in the form of non-thermal
electrons (\cite{hudson-1971ApJ...164..151K};
\cite{hudson-1976SoPh...50..153L}; \cite{hudson-2003ApJ...595L..97H}).
The hard X-ray signature (and hence the energetic dominance of these
electrons) is present whether or not the flare develops the two-ribbon
morphology or has a CME association.

The hard X-ray emission occurs in the impulsive phase of the flare,
contemporaneously with the period of chromospheric evaporation that
fills the coronal loops and with the acceleration phase of the
associated CME (\cite{hudson-2001ApJ...559..452Z}).  In
Section~\ref{hudson-sect:ergs} we describe this phase of the flare
with the thick-target model (\cite{hudson-1971ApJ...164..151K}) which
\citet{hudson-1972SoPh...24..414H} identified with the energy source
of white-light flare continuum.

\section{The Flare Spectrum}\label{hudson-sec:spectrum}

\begin{figure}
\centering
\includegraphics[width=0.55\linewidth]{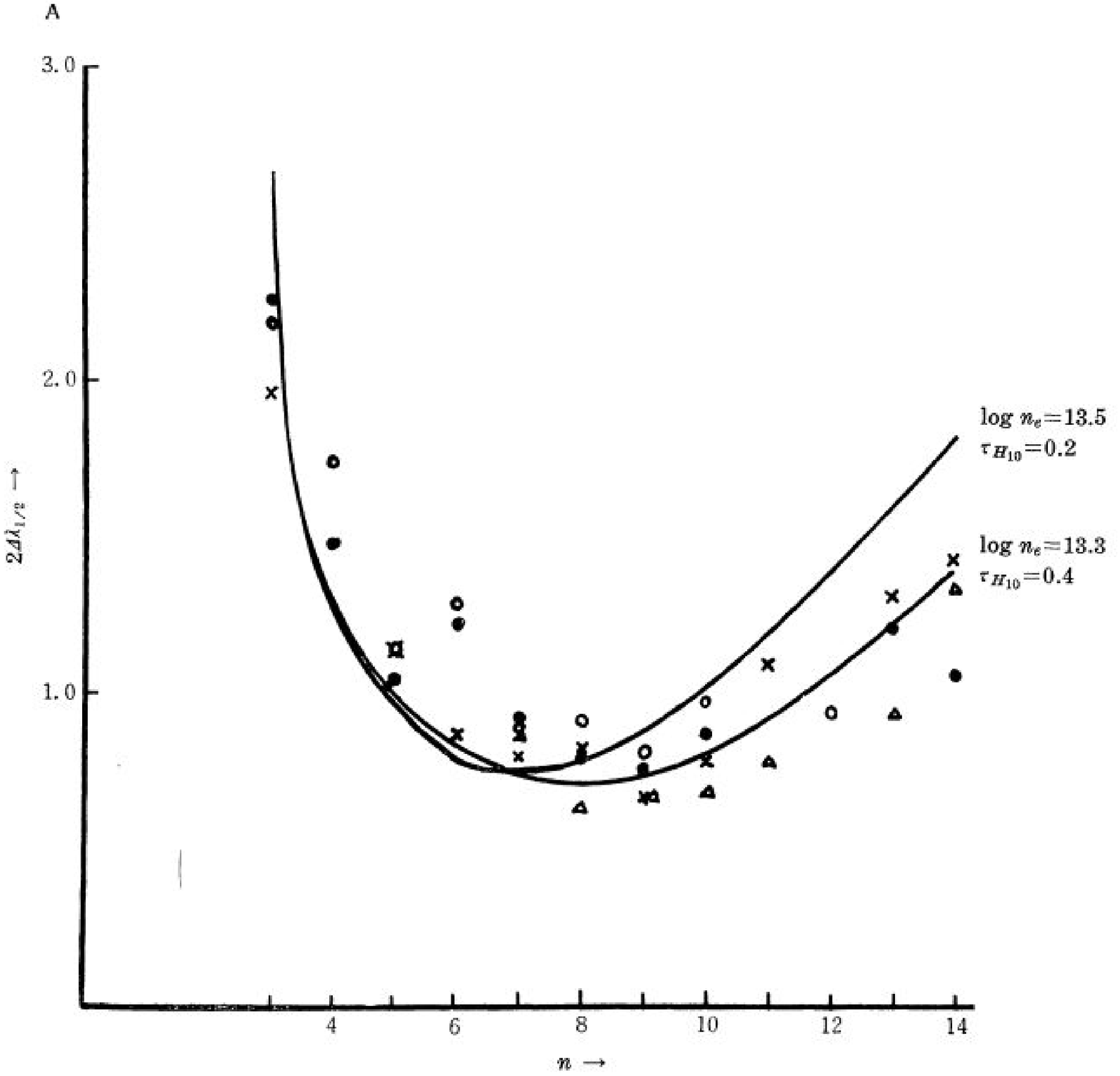}
\includegraphics[width=0.35\linewidth]{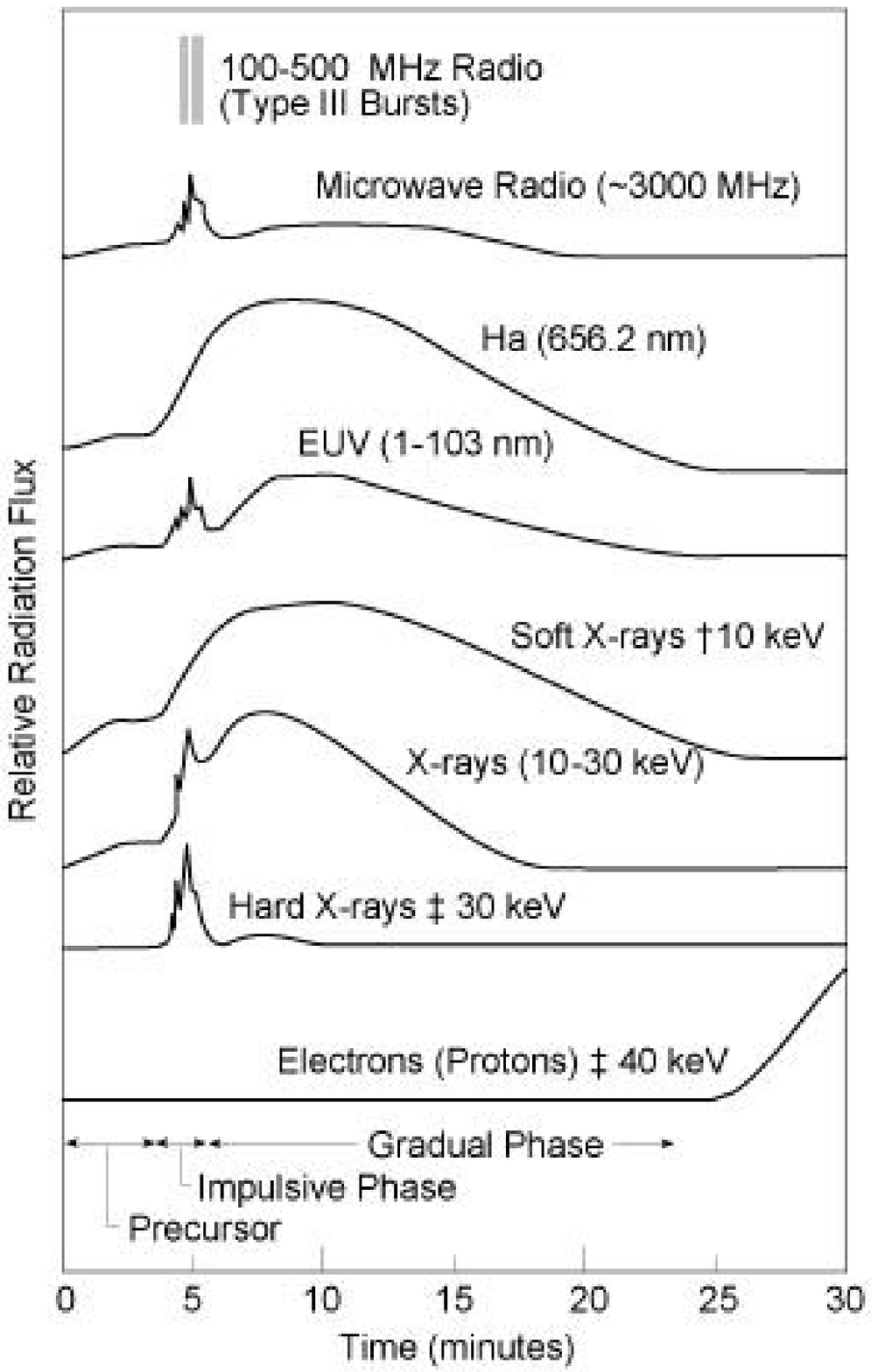}
\caption{{\it Left}: Line widths of the Balmer-series lines, from the
classic paper by \citet{hudson-1959PASJ...11..185S}.  The inferred
densities added to the curves are $\log n_{\rm e} = 13.5$ and 13.3; the
inferred filling factor is small, suggesting either filamentary
structure or thin layering.  {\it Right}: Typical time series of flare
radiations, distinguishing the impulsive phase from the gradual phase
(see \cite{hudson-1971ApJ...164..151K}).  }
\label{hudson-fig:suemoto}
\end{figure}

A (major) flare can be observed at almost any wavelength in a
fast-rise/slow-decay time profile, with some (e.g., the white-light
continuum) having a more impulsive variability, and others (e.g., the
Balmer lines) having a more gradual pattern 
(Figure~\ref{hudson-fig:suemoto}, right).  We generally describe a
flare as consisting of a footpoint and ribbon structures in the lower
atmosphere, coronal loops, and various kinds of ejecta.  The impulsive
phase is typically associated with the footpoint structures, and the
gradual phase with the flare ribbons.  Nowadays imaging spectroscopy
in principle allows us to study these regions independently.

Flare spectroscopy began with the observation of the Balmer series,
which shows broad lines tending towards emission profiles as the flare
gets more energetic.  Early observations of the higher members of the
sequence allowed the inference of a relatively high density and of a
small filling factor (\cite{hudson-1959PASJ...11..185S}; see the left
panel of Figure~\ref{hudson-fig:suemoto}).  Such observations refer to
what we would now call the {\it gradual phase} of the flare (see the
right panel of Figure~\ref{hudson-fig:suemoto} for a sketch of the
temporal development of a flare).  In the {\it impulsive phase} the
continuum appears in emission, as noted originally by Carrington and
by Hodgson independently.  The weak photospheric metallic lines may
also go weakly into emission (or are filled in by continuum) and the
recent observations of \citet{hudson-2004ApJ...607L.131X} show that
flare effects can appear even at the ``opacity minimum'' region of the
spectrum, where one would expect much higher densities.  In fact a
single density could never properly describe such a heterogeneous
structure, but each spectral band provides its own clues.  At the time
of writing no proper analysis of spectroscopic ``response functions''
(e.g., \cite{hudson-2005AGUSMSH12A..02U}) for any of the signatures
has yet been attempted, so our inference of flare structure from the
spectroscopy alone is weak.

The continuum radiation seen in white light and the UV constitutes the
bulk of flare radiated energy (\cite{hudson-1971ApJ...164..151K};
\cite{hudson-2006JGRA..11110S14W}).  TRACE imaging of this emission
component shows it to consist of unresolved, intensely bright fine
structures (\cite{hudson-2006SoPh..234...79H}).  The thick-target
model invokes fast electrons
(energies above about 10~keV) to transport coronal energy into the
chromosphere.  Here collisional losses provide the heating and
footpoint emissions that accompany the hard X-ray bremsstrahlung.  The
thick-target model does not explain the particle acceleration, nor
show how the footpoint sources can be so intermittent.  We return to
this question in Section~\ref{hudson-sec:struct}.

The spectra emitted at the footpoints of the flaring coronal loops
have contributions over an exceptionally broad wavelength range, as
sketched in the right panel of Figure~\ref{hudson-fig:suemoto}.  The
prototypical observable is the hard X-ray flux, which imaging
observations show to be concentrated at the footpoints
(\cite{hudson-1981ApJ...246L.155H}), but impulsive footpoint emissions
also occur in many spectral windows ranging from the microwaves
(limited presumably by opacity) to the $\gamma$-rays (limited
presumably by detection sensitivity).  There is a large body of work
on the H$\alpha$ line alone, both observation and theory.  Berlicki
(2007) reviews the H$\alpha$ spectroscopic material in detail in these
proceedings.  A strong absorption line forms across a wide range of
continuum optical depths, and in principle this single line
might provide sufficient
information to infer the physical structure of the flare everywhere.
In practice the complexities of the radiative transfer and of the
flare motions, especially in the impulsive phase, make this
information ambiguous \citep[see][]{hudson-berlicki}.

\section{Chromospheric Evaporation}\label{hudson-sec:evaporation}

The motions most directly relevant to the chromosphere are often
called ``chromospheric evaporation,'' even though the direct Doppler
signatures of this motion are normally found in lines formed at higher
temperatures (but see \cite{hudson-2005A&A...430..679B}).  That this
process occurs (even if it is not ``evaporation'' strictly speaking)
was suggested by the early observations of loop prominence systems
(e.g., \cite{hudson-1964ApJ...140..746B}) with their ``coronal rain,''
and \citet{hudson-1968ApJ...153L..59N} established its association
with non-thermal processes such as bursts of microwave synchrotron
radiation.  The thermal microwave spectrum (e.g.,
\cite{hudson-1972SoPh...23..155H}) made it particularly clear that the
gradual phase of a solar flare involves the temporary levitation of
chromospheric material into the corona, as opposed to the process that
might be imagined from the earlier term ``sporadic coronal
condensation'' (e.g., \cite{hudson-1963ZA.....56..291W}).  The flows
involved in chromospheric evaporation are along the field direction
and serve to create systems of coronal loops with relatively high gas
pressure and therefore higher (but still probably low) plasma beta.

\begin{figure}
\centering
\includegraphics[width=0.7\linewidth]{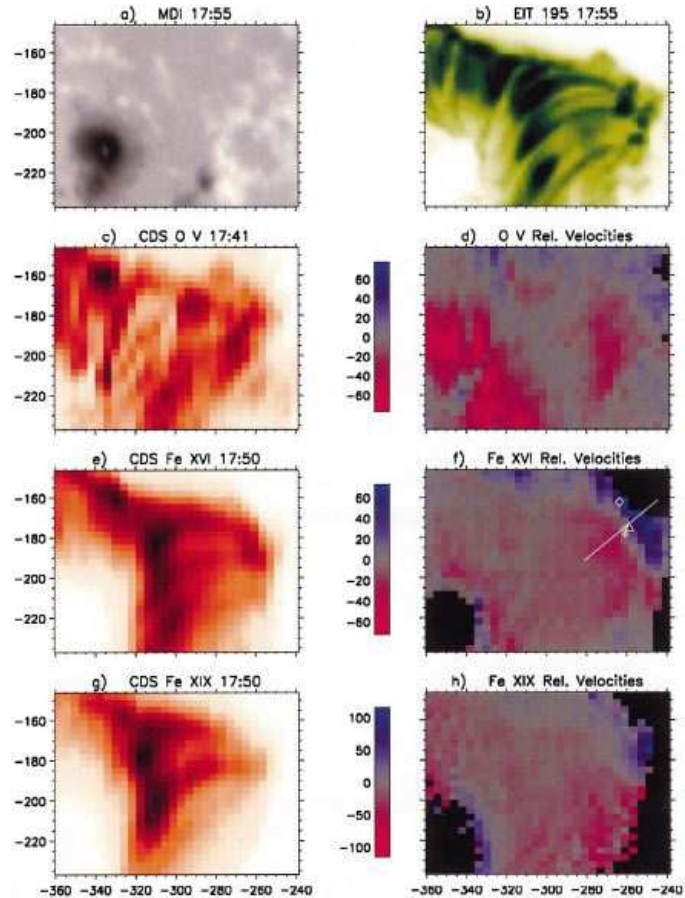}
\caption{Imaging spectroscopy from SOHO/CDS of EUV emission lines in
the gradual phase of a two-ribbon flare, showing the clear signature
of blueshifted upflows in the expected locations along the flare
ribbons.  This is ``gentle'' evaporation not associated with strong
hard X-ray emission (from \cite{hudson-1999ApJ...521L..75C}).  Note
that CDS produces images by scanning in one spatial dimension, so that
each image (while monochromatic) is not instantaneous.  }
\label{hudson-fig:czaykowska}
\end{figure}

The early observational indications of chromospheric evaporation
actually came from blueshifts in EUV and soft X-ray lines (e.g.,
\cite{hudson-1982SoPh...78..107A,hudson-1982ApJ...263..409A}) such as
those from Fe\,XXV or Ca\,XIX.
Figure~\ref{hudson-fig:czaykowska} shows an image-resolved view of
Doppler shifts in an evaporative flow
(\cite{hudson-1999ApJ...521L..75C}).  The chromospheric effects are
more subtle and in fact the impulsive-phase evaporation is difficult
to disentangle from other effects (\cite{hudson-1987ApJ...317..956S}).
The high-temperature blueshifts correspond to upward velocities of
some hundreds km/s and seldom appear in the absence of a stationary
emission line; in other words, hot plasma has already accumulated in
coronal loops as the process continues.  Based on theory and
simulations (\cite{hudson-1985ApJ...289..434F}) one can distinguish
``explosive'' and ``gentle'' evaporation, depending upon the physics
of energy deposition \citep[e.g.,][]{hudson-1999ApJ...521..906A}.  In
explosive evaporation, driven hypothetically by an electron beam, one
has the additional complication of a ``chromospheric condensation''
that produces a redshift as well.  \citet{hudson-1987ApJ...317..956S}
and \citet{hudson-2005A&A...430..679B} survey our overall
understanding.  It would be fair to comment that the explosive
evaporation stage remains ill-understood, even though it principle it
describes the key physics of sudden mass injections into flare loops.

\section{Energetics and Magnetic Field}\label{hudson-sec:ergs}
\label{hudson-sect:ergs}

\begin{figure}
\centering
\includegraphics[width=0.6\linewidth]{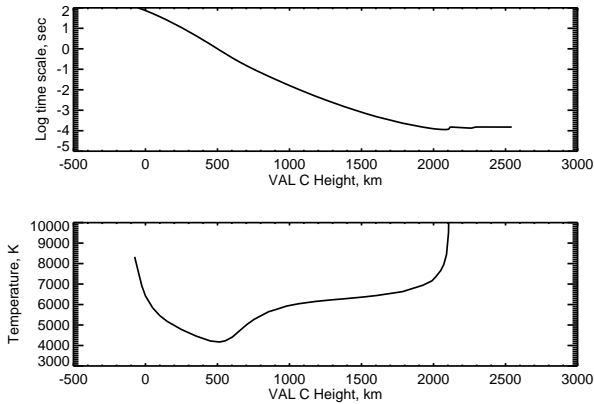}
\caption{Characteristic radiative cooling time (upper) as a function
of height in the VAL-C model, crudely estimated as described in the
text.  The lower panel shows the temperature in this model.  }
\label{hudson-fig:radtime}
\end{figure}

We can use the standard VAL-C model
\citep{hudson-1981ApJS...45..635V}, as discussed further in the
Appendix, to discuss the energetics.  First we establish that the
chromosphere and the rest of the lower solar atmosphere (i.e., that
for which $\tau_{5000} < 1$) have negligible heat capacity and limited
time scales.  Figure~\ref{hudson-fig:radtime} shows an estimate of the
radiative time scale in the VAL-C model
(\cite{hudson-1981ApJS...45..635V}).  This shows 3$\sigma(z)kT /
{\mathcal L}_{\odot}$, where $\sigma$ is the surface density as a
function of height about the $\tau_{5000}$~=~1 layer, and ${\mathcal
L}_{\odot}$ the solar luminosity.  The time scale decreases below
1~sec only above z~$\sim$~515~km, near the temperature minimum in the
VAL-C model.  Above this height any energy injected into the system
will tend to radiate rapidly, resulting in a direct energy balance
between input and output energy, rather than a local storage and
release.  At lower altitudes we would not expect to see rapid
variability.

The model also allows us to ask whether the chromosphere itself can
store energy comparable to that released in a major flare or CME.
Table~\ref{hudson-tab:chrom} gives some order-of-magnitude properties
for a chromospheric area of 10$^{19}$~cm$^{2}$, showing both possible
sources (bold) and sinks (italics) of energy.  For the magnetic field
we simply assume 10~or 1000~G as representative cases.  Using the
total magnetic energy in this manner is an upper limit, since the
actual free energy would depend on its degree of non-potentiality.  
We find that
magnetic energy storage limited to the volume of the chromosphere will
not suffice, unless unobservably small-scale fields there somehow
dominate.  The gravitational potential energy also will not suffice.
Estimates of this sort confirm the idea that the flare energy must
reside in the corona prior to the event.

The estimate of gravitational potential energy is somewhat more
ambiguous.  The Table shows the value needed to displace the entire
atmosphere by its total thickness, the equivalent of roughly 3$''$ in
the VAL-C model.  There does not seem to be any evidence for such a
displacement, although I am not aware of any searches.  It is likely
that the stresses that store energy in the coronal field have their
origin deeper in the convection zone, rather than in the atmosphere
\citep{hudson-1989sspp.conf..219M}.
Actually the observable changes of gravitational energy are even of
the wrong sign, given that we normally observe only outward motions,
(against gravity) during a flare.


\begin{table}
\caption{Properties of a chromospheric volume of area 10$^{19}$~cm$^{2}$}\label{hudson-tab:chrom}
\begin{tabular*}{\textwidth}{@{\extracolsep{\fill}}lll}
  \hline\cr
  Parameter & VAL-C &VAL-C above T$_{min}$ \cr
  \hline
  Mass & 4$\times$10$^{19}$ gram & 5$\times$10$^{17}$ gram \cr
  {\bf Magnetic energy} &  1$\times$10$^{28} $erg & 8$\times$10$^{27}$ erg  (10 G) \cr
  {\bf Magnetic energy} &  1$\times$10$^{32}$ erg& 8$\times$10$^{31}$ erg  (10$^{3}$ G) \cr
  {\bf Gravitational energy}$^{a}$ &  3$\times$10$^{32}$ erg &  3$\times$10$^{30}$ erg \cr
  {\it Thermal energy} & 2$\times$10$^{31}$ erg &  3$\times$10$^{29}$ erg \cr
  {\it Kinetic energy} & 3$\times$10$^{29}$ erg &  3$\times$10$^{27}$ erg \cr
   {\it Ionization energy} & 4$\times$10$^{32}$ erg &  6$\times$10$^{30}$ erg \cr
 \hline
\end{tabular*}
$^{a}$Potential energy for a vertical displacement of 2.5~$\times$~10$^{8}$ cm
\end{table}

From Table~\ref{hudson-tab:chrom} one concludes that the chromosphere
probably does play a dominant role in the energetics of a solar flare,
at least as described by a semi-empirical model such as VAL-C.  This
just restates the conventional wisdom, namely that the flare energy
needs to be stored magnetically in the corona, rather than in the
chromosphere where the radiation forms.  Note that this is backwards
from the relationship for steady emissions: the requirement for
chromospheric heating is {\it larger} than that for coronal heating,
so it is possible to argue that the steady-state corona actually forms
as a result of energy leakage from the process of chromospheric
heating (e.g., \cite{hudson-1994ApJ...427..446S}).

We can make a similar estimate for energy flowing up from the
photosphere.  The Alfv{\'e}n speed at $\tau_{5000}$~=~1 ranges from
3~to~30 km~s$^{-1}$, depending on the field strength, in the VAL-C
model (see Appendix).  Below the surface of the Sun $v_{A}$ drops
rapidly because of the increase of hydrogen ionization.  Thus
chromospheric flare energy cannot have been stored just below the
photosphere, since it could not propagate upwards rapidly enough
\citep{hudson-1989sspp.conf..219M}.
This again supports the conventional wisdom that flare energy resides
in the corona prior to the event.

To drive chromospheric radiations from coronal energy sources requires
efficient energy transport, which is normally thought to be in the
form of non-thermal particles \citep[the ``thick target
model'',][]{hudson-1971SoPh...18..489B,hudson-1972SoPh...24..414H} or
in the form of thermal conduction as in the formation of the classical
transition region.  Both of these mechanisms provide interesting
physical problems, but the impulsive phase of the flare (where the
thick-target model usually is thought to apply) certainly remains less
understood.  Section~\ref{hudson-sec:model} comments on models.

The magnetic field in the chromosphere is decisively important but
ill-understood.  The plasma beta is generally low (see Appendix), so
just as in the corona the dynamics depends more on the behavior of the
field itself than to the other forces at work.  Generally we believe
that the subphotospheric field exists in fibrils, implying the existence
of sheath currents that isolate the flux tubes from their unmagnetized
environment.  On the other hand, the dominance of plasma pressure in
the chromosphere as well as the corona implies that the field must
rapidly expand to become space-filling.
\citet{hudson-2000ApJ...545.1089L} discuss the physics involved in
this process as flux emerges from the interior.  The effect of the
flux emergence must be to create current systems linking the sources
of magnetic stress below the photosphere, with the non-potential
fields containing the coronal free energy.  A full theory of how this
works does not exist, and we must add to the uncertainty the
possibilty of unresolved fields
\citep[e.g.,][]{hudson-2004Natur.430..326T}.  Their suggested factor
of 100 in {\bf B}$^{2}$ would clearly affect the estimate of magnetic
energy given in Table~\ref{hudson-tab:chrom} and perhaps change 
everything.
We note in this context that the ``impulse response" flares
\citep{hudson-1992ApJ...384..656W} have scales so small that one
could argue for an entirely chromospheric origin.

\section{Energetics and the Formation of the Continuum}\label{hudson-sec:cont}

The formation of the optical/UV emission spectrum of a solar flare has
from the outset presented a special challenge, since (a) it represents
so much energy, and (b) it appears in what should be the stablest layer
of the solar atmosphere.  The recent observations of rapid variability
and spatial intermittency make this all the more interesting, and
these observations -- now from space -- also help to intercompare
events; previous catalogs of white-light flares (e.g.,
\cite{hudson-1989SoPh..121..261N} and references therein) had to be
based on spotty observations made with a wide variety of instruments.
Observationally, the continuum appears to have two classes, with most
events (``Type I'' spectra) showing evidence for recombination
radiation via the presence of the Balmer edge and sometimes the
Paschen edge as well.  A few events (e.g.,
\cite{hudson-1974SoPh...38..499M}) show spectra with weak or
unobservable Balmer jumps, implicating H$^{-}$ continuum as observed
in normal photospheric radiation.  The spectra in the latter class
(``Type II'') suggests a relationship to Ellerman bombs
(\cite{hudson-2001ChJAA...1..176C}).  However, the physics of Ellerman
bombs appears to be quite different from that of solar flares
\citep[e.g.,][]{hudson-2004ApJ...614.1099P}, though.

The strong suggestion from correlations is that non-thermal electrons
physically transport flare energy from the corona, where it had been
stored in the current systems of non-potential field structures, into
the radiating layers.  The hard X-ray brems\-strahlung results from
the collisional energy losses of these particles, and other signatures
(such as the optical/UV continuum) depends on secondary effects.
Proposed mechanisms include direct heating, heating in the presence of
non-thermal ionization, and radiative backwarming.  In some manner
these effects (or others not imagined) must provide the emissivity
$\epsilon_{\nu}$, to support the observed spectrum.  Note that the
emissivity is often expressed in terms of the source function
$S_{\nu}$~=~$\epsilon_{\nu}/\kappa_{\nu}$ via the opacity
$\kappa_{\nu}$.  In a steady state one would have energy balance
between the input (e.g., electrons) and the continuum.  Fletcher et
al. (2007) have now shown that this implies energy transport by
low-energy electrons, below 25~keV, as opposed to the 50~keV or higher
suggested by some earlier authors.  Such low-energy electrons have
little penetrating power and could not directly heat the photosphere
itself from a coronal acceleration site.  Thus either the continuum
arises from altered conditions in the chromosphere, or some mechanism
must be devised to link the chromosphere and photosphere not involving
the thick-target electrons.

``Radiative backwarming'' -- for example Balmer and Paschen continuum
excited in the chromosphere and then penetrating down to and heating a
deeper layer -- could in principle provide a vertical step between
energy source and sink.  One problem with this is that the weaker
backwarming energy fluxes might not cause appreciable heating in the
denser atmosphere, and thus not be able to contribute to the observed
continuum excess, because of the short radiative time scale.  This
idea is a variant of the mechanism of non-thermal ionization
originally proposed by \citet{hudson-1972SoPh...24..414H} in the
``specific ionization approximation,'' which involves no
radiative-transfer theory and simply assumes ion-electron pairs to be
created locally at a mean energy ($\sim$30~eV per ion pair).  Finally,
the rapid variability observed in the continuuum, even at 1.56~$\mu$m
\citep{hudson-2006ApJ...641.1210X} provides a clear argument that the
continuum forms at the temperature minimum or above (see
Section~\ref{hudson-sec:ergs}, especially
Figure~\ref{hudson-fig:radtime}).

Early proponents of particle heating as an explanation for white-light
flares also considered protons as an energy source
(\cite{hudson-1970SoPh...15..176N,hudson-1970SoPh...13..471S}).  This
made sense, because protons at energies even below those
characteristic of $\gamma$-ray emission-line excitation can penetrate
more deeply than the electrons that produce hard X~rays.  It makes
even more sense now that we have the suggestion that ion acceleration
in solar flares may rival electron acceleration energetically
(\cite{hudson-1995ApJ...455L.193R}).
\citet{hudson-1990SoPh..130..253S} suggest that particle acceleration
in solar flares involves a neutral beam, implying that the major
energy content (and hence the optical/UV continuum) would originate in
the ion component.  This idea does not appear to explain the apparent
simultaneity of the footpoint sources
(\cite{hudson-1996AdSpR..17...67S}), and at present we do not
understand the plasma physics of the particle acceleration and
propagation well enough even to identify the location of the
acceleration region.

\section{Flare Structures Inferred from Chromospheric Signatures}\label{hudson-sec:struct}

\begin{figure}
\centering
\includegraphics[width=0.8\linewidth]{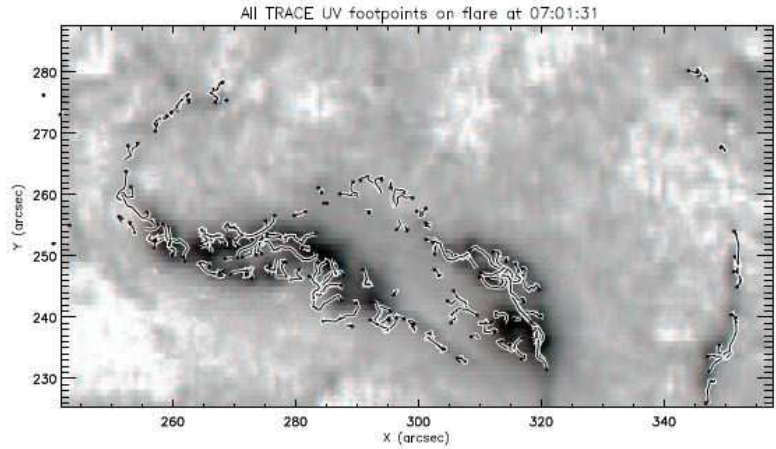}
\caption{Flare footpoint apparent motions deduced from TRACE UV observations.
Each squiggle represents the track of a bright point visible for several consecutive images at a
few-second cadence, with the black dot showing the beginning of each track \citep{hudson-2004SoPh..222..279F}.
}
\label{hudson-fig:pollock}
\end{figure}

\begin{figure}
\centering
\includegraphics[width=0.7\linewidth]{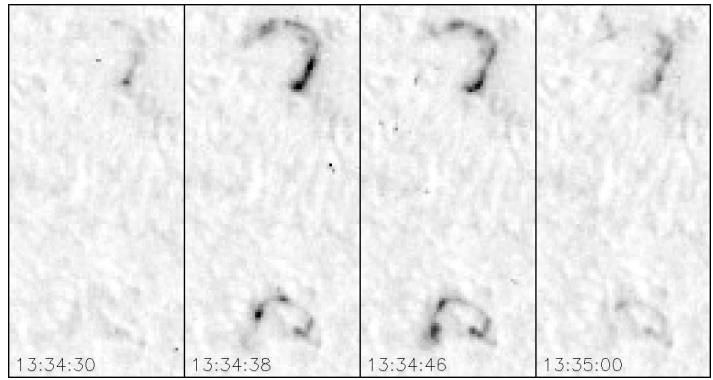}
\caption{Intermittent structure seen in TRACE white-light images of an M-class flare
on July 24, 2004.
The individual frames have dimensions 32$''\times\ $64$''$.
Note the presence of bright features consistent with the TRACE angular resolution, and which change from frame to frame over the 30-second interval.
These observations do not appear to resolve the fluctuations either in space or in time
\citep{hudson-2006SoPh..234...79H}.
}
\label{hudson-fig:4panel}
\end{figure}

The continuum kernels may move systematically for perhaps tens of
seconds and generally have short lifetimes.  We illustrate this in
Figure~\ref{hudson-fig:pollock} (from
\cite{hudson-2004SoPh..222..279F}).  This shows the motions of
individual UV~bright points within the flare ribbon structure.  Such
motions are only apparent motions, as in a deflagration wave, because
they exceed the estimated photospheric Alfv{\' e}n speed (see
Section~\ref{hudson-sec:evaporation} and the Appendix).
Figure~\ref{hudson-fig:4panel} (from
\cite{hudson-2006SoPh..234...79H}) makes the same point for a
different flare, using TRACE white-light observations.  The basic
picture one gets from such observations is that the white light/UV
continuum of a flare appears in compact structures that are
essentially unresolved in space and in time within the present
observational limits.  These bright points contain enormous energy and
thus must map directly to the energy source.  We do not know if the
fragmentation (intermittency) results from this mapping or is
intrinsic to the basic energy-release mechanism.

How do the small chromospheric sources map into the corona, where the
flare energy must reside on a large scale before its release?  A
strong literature has grown up regarding this point, interpreting the
ribbon motions as measures of flux transfer in the standard
magnetic-reconnection model (\cite{hudson-1986lasf.symp..453P}; see
also literature cited, for example, by
\cite{hudson-2005ApJ...632.1184I}).  The flux transfer in the
photosphere is taken to measure the coronal inflow into the
reconnecting current sheet, which appears to correlate with the
radiated energy as seen in hard X-rays, UV, or H$\alpha$.
Figure~\ref{hudson-fig:bruzek} (right) shows the assumed geometry
linking the chromosphere and corona.  The analysis extends to the
multiple simultaneous UV footpoints apparently moving within the
ribbons as they evolve, as noted in Figure~\ref{hudson-fig:pollock}
above.  The analyses suggest a strong relationship between energy
release and the inferred coronal Alfv{\' e}n speed.

\section{Dynamics and Magnetic Reconnection}\label{hudson-sec:recon}

To release energy from coronal magnetic field in a largely
``frozen-field'' plasma, a flare must involve mass motions.  We often
do observe apparent motions, both parallel and perpendicular to the
field as indicated by the image striations (``loops'').  Most of the
observable motions are outward, leading to the idea of a ``magnetic
explosion'' (e.g.,  \cite{hudson-2001ApJ...552..833M}).  Motions
apparently perpendicular to the magnetic field may become coronal mass
ejections (CMEs) and contain a great deal of energy (e.g.,
\cite{hudson-2005JGRA..11011103E}).  These perpendicular motions also
are involved in flare energy release; for example the large-scale
magnetic reconnection involved in many flare models
(Figure~\ref{hudson-fig:bruzek}, right panel) necessarily involve
``shrinkage'' (e.g., \cite{hudson-1987SoPh..108..237S};
\cite{hudson-1996ApJ...459..330F}).
Note that this process is more of a magnetic implosion than a magnetic
explosion \citep{hudson-2000ApJ...531L..75H}.

The motion of flare footpoints and ribbons is (we believe) only
apparent, because of the low Alfv{\' e}n speed $v_{A}$ in the
photosphere, where the field is temporarily anchored (``line-tied'').
For $B$~=1000~G and $n$~=~10$^{17}$~cm$^{-3}$ we find $v_{A}
\sim$~6~km~s$^{-1}$; observations often suggest motions an order of
magnitude faster \citep[e.g.,][]{hudson-2006ApJ...650.1184S}.  The
motions therefore represent a wave-light conflagration moving through
a relatively fixed magnetic-field structure.  It is natural to imagine
that this sequence of field lines links to the coronal energy-release
site, which the standard model identifies with a current sheet that
mediates large-scale magnetic reconnection.

\begin{figure}
\centering
\includegraphics[width=0.46\linewidth]{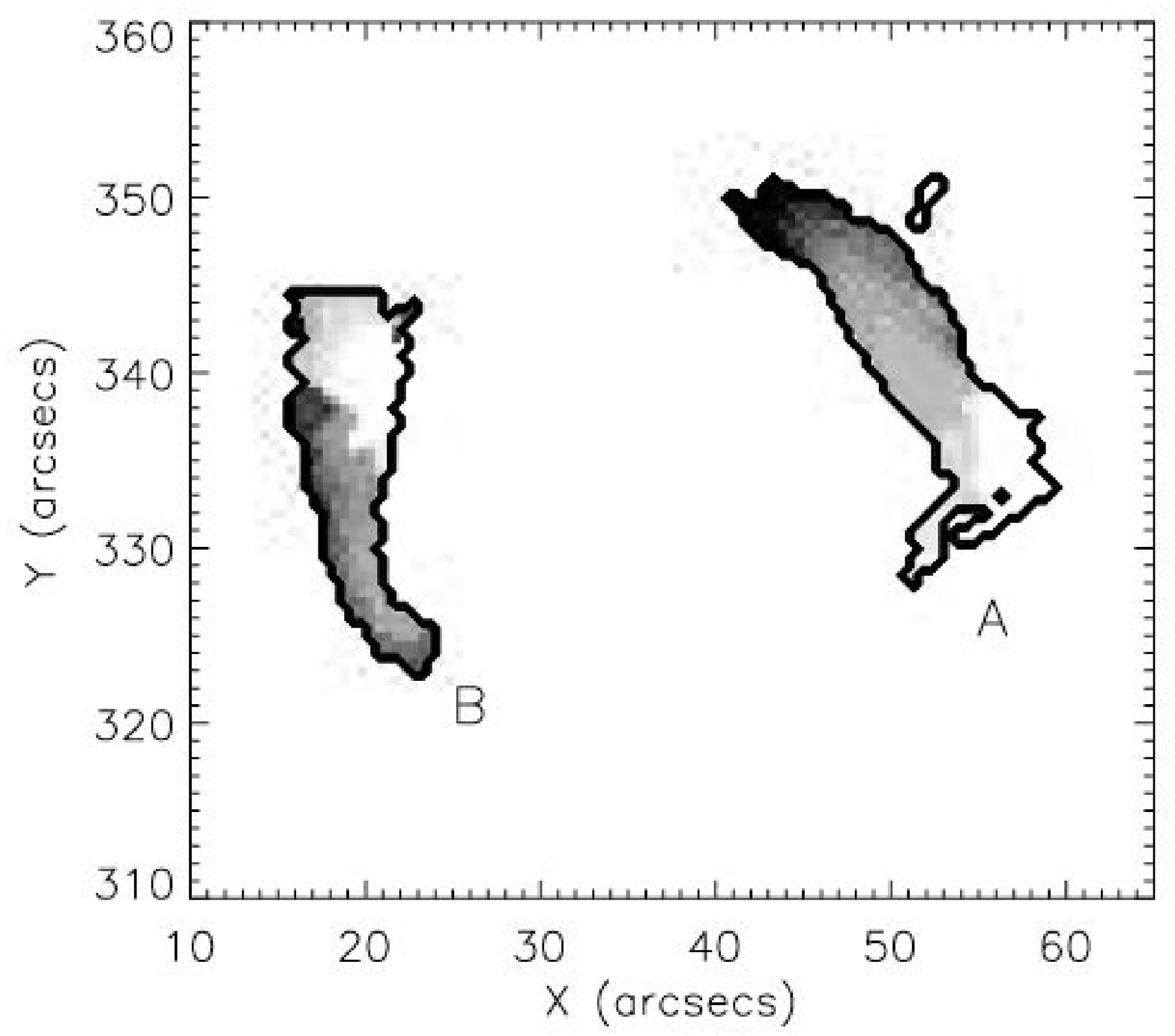}
\includegraphics[width=0.42\linewidth]{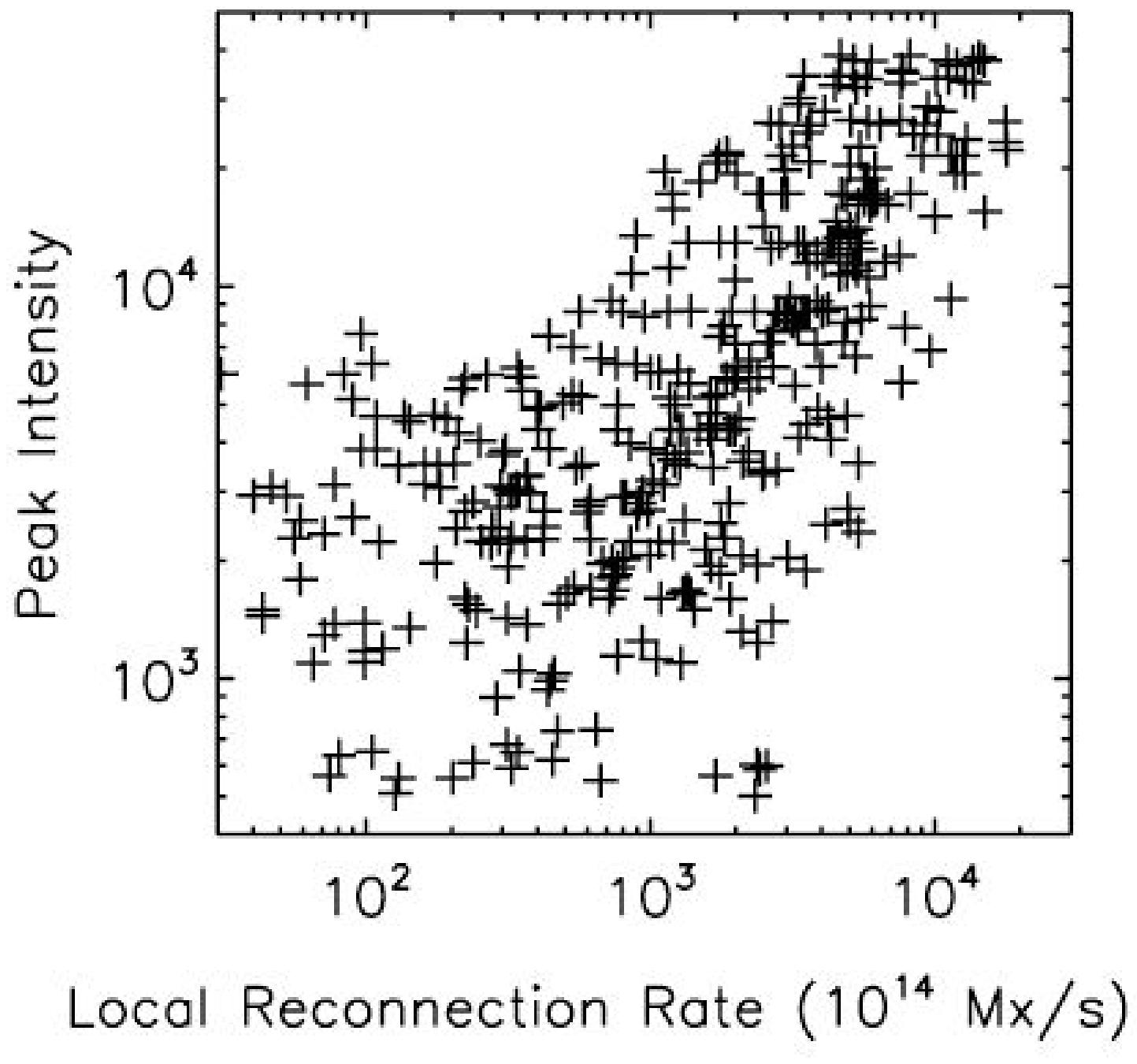}
\caption{Left: UV ribbons (TRACE observations) from a flare of November 23, 2000.
The gray scale shows the time sequence of brightening.
Right: Correlation between pixel brightness in Ribbon~A and the inferred reconnection rate
(from \cite{hudson-2006ApJ...641.1197S}).
}
\label{hudson-fig:saba}
\end{figure}

Figure~\ref{hudson-fig:saba} shows one example of the result of an
analysis of the apparent motion of a flare ribbon
\citep{hudson-2006ApJ...641.1197S}.  This and other similar analyses
reveal a tendency for the ``reconnection rate'' to correlate with the
pixel brightness.  The reconnection rate is the rate at which flux is
swept out in the ribbon motion, often expressed as an electric field
from {\bf E}~=~{\bf v}$\times${\bf B} (the so-called ``reconnection
electric field'').  In this picture the flare ribbons are identified
with ``quasi-separatrix structures'' where magnetic reconnection can
take place most directly.

\section{Surges, Sprays, and Jets}\label{hudson-sec:surge}

Chromospheric material also appears in the corona in the form of
surges and sprays, which may have a close relationship to the flare
process \citep[e.g.,][]{hudson-1980IAUS...91..173E}.  In addition, of
course, we observe filaments and prominences in chromospheric lines,
and these also have a flare/CME association, but too tangential for
discussion in this review.

Surges and sprays are H$\alpha$ ejecta, rising into the corona as a
result of chromospheric magnetic activity.  The literature
traditionally distinguishes them by apparent velocity, with the
faster-moving sprays taken to have stronger flare associations.
Surges often appear to return to the Sun, while sprays accelerate
beyond the escape velocity and do not return.  Both appear to move
along the magnetic field lines, but unlike the evaporation flow the
surges and sprays incorporate material at chromospheric temperatures.

\begin{figure}
\centering
\includegraphics[width=0.7\linewidth]{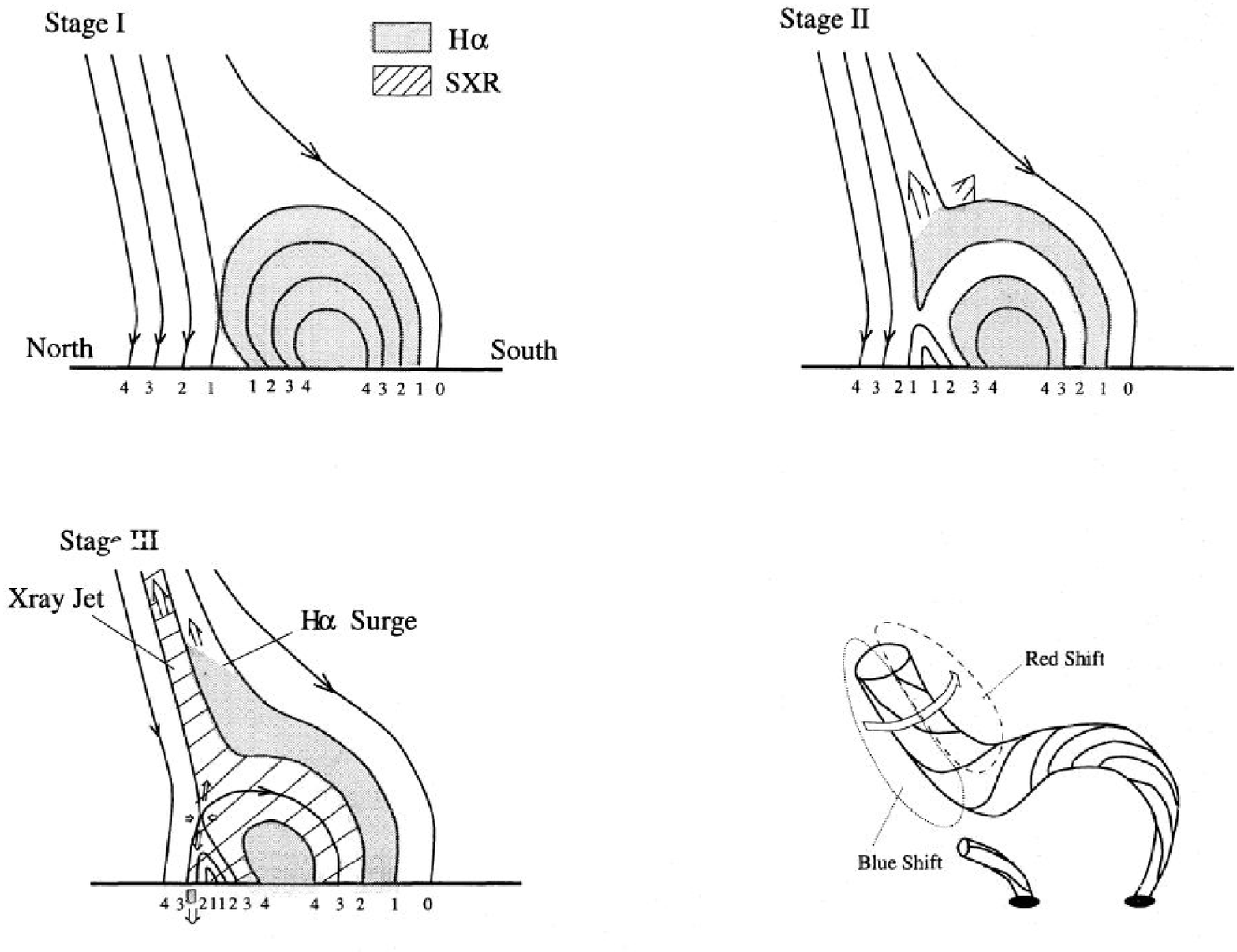}
\caption{A mechanism for jet/surge formation involving emerging flux
(upper left), with magnetic reconnection against already-open fields
(upper right), which may lead to a high-temperature ejection (the jet)
entraining chromospheric material (the surge).  The cartoon at lower
right describes the observations of
\citep{hudson-1996ApJ...464.1016C}, who find a spinning motion
suggesting that the process must occur in a 3D configuration rather
than that of the cartoons left and above.  }
\label{hudson-fig:canfield}
\end{figure}

Modern soft X-ray and EUV data ({\it Yohkoh}, SOHO, and TRACE) have
had sufficient time resolution to reveal the phenomenon of X-ray
jets \citep{hudson-1992PASJ...44L.173S}; 
see also the UV observations of \cite{hudson-1983ApJ...267L..65D}.  
These tightly-collimated
structures at X-ray temperatures have a strong correlation with surges
and sprays, and indeed presumably lead to the jet-like CMEs seen at
much greater altitudes \citep{hudson-2002ApJ...575..542W}.  These
events have a strong association with emerging flux, and indeed the
X-ray jets invariably have an association with microflares and
originate in the chromosphere near the microflare loop(s)
\citep{hudson-1992PASJ...44L.173S}.  As Zirin famously remarked, most
emerging flux emerges within active regions, and that is where the
jets preferentially occur.  The site is frequently in the leading part
of the sunspot group.  Figure~\ref{hudson-fig:canfield}
\citep{hudson-1996ApJ...464.1016C} shows the sequence of events in an
explanation of these phenomena invoking magnetic reconnection to allow
chromospheric material access to open fields.  Note that this scenario
imposes two requirements on the chromosphere: there must be open and
closed fields juxtaposed, and a large-scale reconnection process must
be able to proceed under chromospheric conditions.  The
\citet{hudson-1996ApJ...464.1016C} observations strongly imply that
this process requires the presence of vertical electric currents
supporting the observed twisting motions.

The surges, sprays, and jets, not to mention flares and CMEs,
underscore the time dependence and three-dimensionality characterizing
what is often characterized as a thin time-independent layer for
convenience.  The subject of spicules is outside the scope of this
review, but we note that they represent a form of activity that occurs
ubiquitously outside the magnetic active regions.

\section{A Chromospheric $\gamma$-ray Mystery}\label{hudson-sec:gamma}

\begin{figure}
\centering
\includegraphics[width=0.65\linewidth]{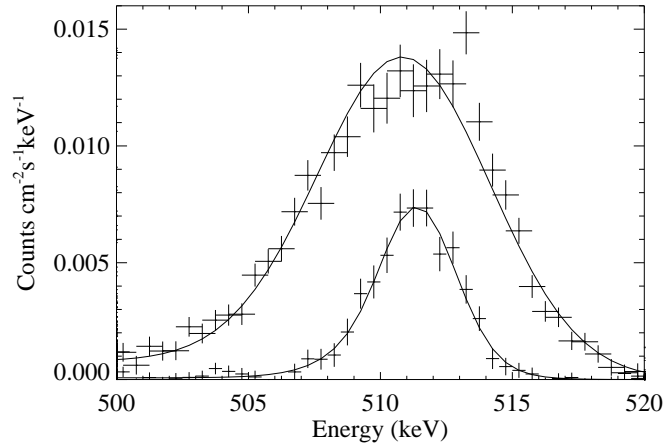}
\caption{RHESSI $\gamma$-ray observations of the 0.511~MeV line of
positron annihilation (\cite{hudson-2004ApJ...615L.169S}).  The two
line profiles are from different integrations in the late phase of the
X17~flare of 28~October 2003; for the broader line the authors suggest
thermal broadening, which would require a large column depth of
transition-region temperatures during the flare.  }
\label{hudson-fig:share}
\end{figure}

The $\gamma$-ray observations of solar flares have begun, as did the
radio and X-ray observations before them, to open new windows on flare
physics.  \citet{hudson-2004ApJ...615L.169S} have made a discovery
that is difficult to understand and which involves chromospheric
material.  They report observations of the line width of the 0.511~MeV
$\gamma$-ray emission line formed by positron annihilation
(Figure~\ref{hudson-fig:share}).  This emission requires a complicated
chain of events: the acceleration of high-energy ions, their
collisional braking and nuclear interactions in the solar atmosphere,
the emission of secondary positrons by the excited nuclei, the
collisional braking of these energetic positrons in turn, and finally
their recombination with ambient electrons to produce the 0.511~MeV
$\gamma$-rays.  Because the $\gamma$-ray observations are so
insensitive, this process requires an energetically significant level
of particle acceleration that is possibly distinct from the well-known
electron acceleration in the impulsive phase.

The mystery comes in the line width of the emission line.
Surprisingly the pioneering RHESSI observations of
\citet{hudson-2004ApJ...615L.169S} showed it to be broad enough to
resolve.  The likeliest source of this line broadening is Doppler
motions in the positron-annihilation region.  This requires the
existence of a large column density (of order gram~cm$^{-2}$) at
transition-region temperatures; the transition region under
hydrostatic conditions would be many orders of magnitude thinner (see
also Figure~\ref{hudson-fig:ztausigma}).  According to
\citet{hudson-2006ApJ...650.1184S}, the excitation of the footpoint
regions during the the time of intense particle acceleration only
continues for some tens of seconds at most.  This would represent the
time scale for the apparent motion of a foopoint source across its
diameter.  The $\gamma$-ray observations, on the other hand, require
minutes of integration for a statistically significant line-profile
measurement.

We therefore are confronted with a major problem.  What is the
structure of the flaring atmosphere that permits the formation of the
broad 0.511~MeV $\gamma$-ray line?
Recent spectroscopic observations of the impulsive phase in the~UV, as viewed
off the limb \citep{hudson-2007astro.ph..1359R} make a conventional
explanation difficult.

\section{Theory and Modeling}\label{hudson-sec:model}

To understand the chromospheric spectrum of a solar flare we must
understand the formation of the radiation and its transfer in the
context of the motions produced by (or producing) the flare.  The
representation of the spectrum by a ``semi-empirical model''
represents one shortcut; in such an approach (e.g., the
standard VAL model that we use in the Appendix) one attempts
to construct a model atmosphere capable of describing the spectrum
even if it may not be physically self-consistent.  Such descriptions
may however be sufficient in the gradual phase of a flare when the
flare loops no longer have energy input and simply evolve by cooling
and draining.  Even here, however, we do not have a good understanding
of the ``moss'' regions that form at the footpoints of these
high-pressure loops (but see \cite{hudson-1999SoPh..190..409B}).  So
far as I am aware there is no literature specifically on ``spreading
moss,'' the similar structure that appears in association with flare
ribbons.

A more complete approach to the physics comes from ``radiation
hydrodynamics'' physical models, most recently those of
\citet{hudson-2005ApJ...630..573A}; see \citet{hudson-berlicki}  for a fuller
description.  Such models solve the equations of hydrodynamics and
radiative transfer simultaneously and can thus deal with chromospheric
evaporation and the formation of the high-pressure flare loops.  This
framework is necessary if we are to be able to understand the flare
impulsive phase \citep[e.g.,][]{hudson-2003AdSpR..32.2393H}.  Even
these models do not have sufficient realism, though, since they work
currently in one dimension and thus cannot follow the time development
of the excitation properly; the high-resolution observations of UV and
white light by TRACE clearly show that the energy release has
unresolved scales.  Further, as pointed out by
\citet{hudson-1972SoPh...24..414H}, the ionization of the chromosphere
(and hence the formation of the continua) cannot be described by a
fluid approximation, or even by non-LTE radiative transfer that
assumes a unique temperature.

\begin{figure}
\centering
\includegraphics[width=0.95\linewidth]{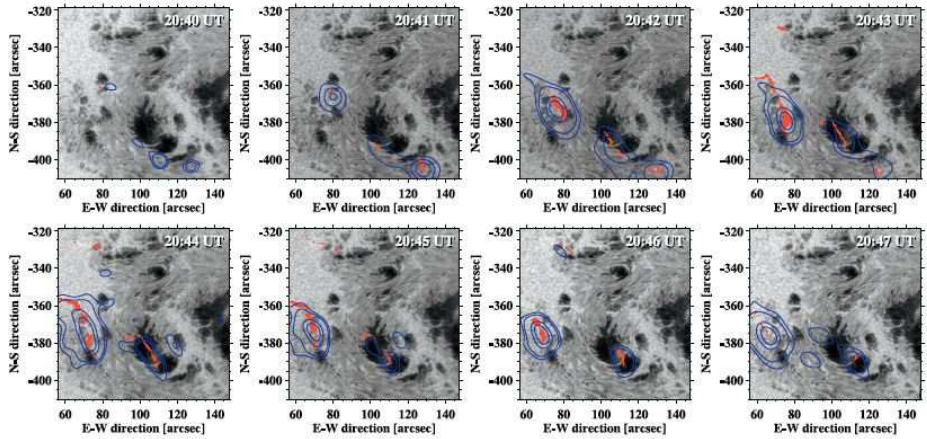}
\caption{Continuum emission in the near infrared (1.56~$\mu$m, the
``opacity minimum'' region) during an X10 flare
(\cite{hudson-2004ApJ...607L.131X}).  Red shows the IR emissions,
contours show the RHESSI 50-100~keV X-ray sources.  The IR contrast
relative to the preflare photosphere reached $\sim$20\% in this event.
}
\label{hudson-fig:xu}
\end{figure}

At present there has been little effort to create an electrodynamic
theory of chromospheric flare processes, even though non-thermal
particles are widely thought to provide the dominant energy in at
least the impulsive phase.  In the gradual phase there is interesting
physics associated with heat conduction because the transition region
would have to become so steep that classical conductivity estimates
have difficulty (\cite{hudson-1983ApJ...266..339S}).  A more complete
theory would have to take plasma effects into account and would
probably contain elements of theories of the terrestrial aurora that
are now largely missing from the solar lexicon.  This lack of
self-consistency in the modeling probably means that we have major
gaps in our understanding of, for example, the evaporation process as
it affects the fractionation of the elements and of the ionization
states of the flare plasma.  The Appendix gives estimates of the
ranges of some the key plasma parameters in the chromosphere.

\section{Conclusions}

This article has reviewed chromospheric flare observations from the
point of view of the newest available information -- {\it Yohkoh},
SOHO, TRACE and RHESSI, for example, but not {\it Hinode} or STEREO
(already launched), nor much less FASR or ATST (not launched yet at
the time of writing).
 spite of the high quality of the data prior to these missions, we
still find major unsolved problems:

\bigskip\noindent  
$\bullet$ How does the chromosphere obtain all of the energy that it radiates?

\bigskip\noindent
$\bullet$ How can flare effects appear at great depths in the photosphere?

\bigskip\noindent
$\bullet$ How is the anomalous 0.511~MeV line width produced?

\bigskip\noindent
$\bullet$ What are the elements of an electrodynamic theory of chromospheric flares?

\bigskip In my view the solution of these problems cannot be found in
chromospheric observations alone, because the physical processes
involve much broader regions of the solar atmosphere.  Even providing
answers to these specific questions may not reveal the plasma physics
responsible for flare occurrence, which may involve spatial scales too
fine ever to resolve.  But we can hope that new observations from
space and from the ground, in wavelengths ranging from the radio to
the $\gamma$-rays, will enable us to continue our current rapid
progress, and can speculate that eventually numerical tools will
supplement the theory well enough for us to achieve full comprehension
of the important properties of flares.  To get to this point we will
need to deal with the chromosphere, as messy as it is.

One important task that is probably within our grasp now is the
computation of response functions for physical models of flares.  At
present these are restricted to very limited numerical explorations of
the radiative transfer within the framework of one-dimensional
radiation hydrodynamics (e.g., \cite{hudson-2005ApJ...630..573A}).
The energy transport in these models has been restricted to simplistic
representations of particle beams for energy transport, and do not
take account of complicated flare geometries, waves, or various
elements of plasma physics.  Future developments of chromospheric
flare theory will need to complete the picture in a more
self-consistent manner.

\bigskip
\noindent{\bf Acknowledgments} This work has been supported by NASA
under grant NAG5-12878 and contract NAS5-38099.  I thank W. Abbett for
a critical reading.  I am also grateful to Rob Rutten for LaTeX
instruction, and to Bart de Pontieu for meticulous keyboard entry.
 
\bigskip\bigskip\noindent {\bf Appendix: plasma
parameters}\label{hudson-sec:appendix}

\begin{figure}
\centering
\includegraphics[width=0.75\linewidth]{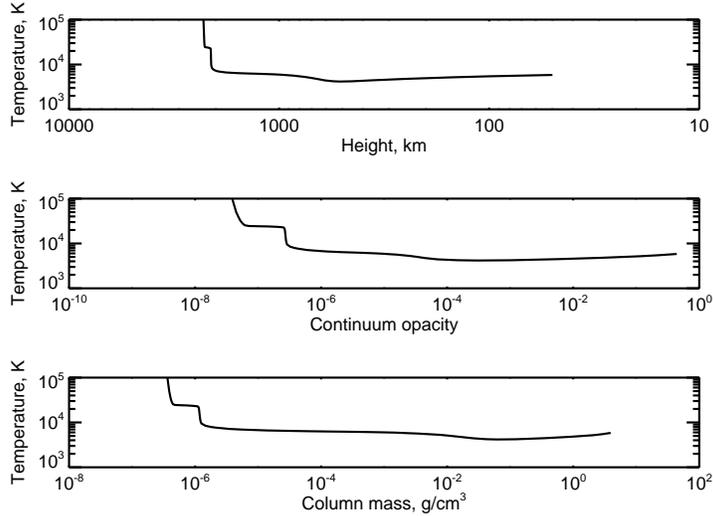}
\caption{Illustration of the structure of a semi-empirical model, using three different independent
variables: the VAL-C temperature plotted against height, optical depth, and column mass.
}
\label{hudson-fig:ztausigma}
\end{figure}

\bigskip\noindent The lower solar atmosphere marks the transition
layer between regions of striking physical differences, and as one
goes further up in height the tools of plasma physics should become
more important.  This Appendix evaluates for convenience several basic
plasma-physics parameters for the conditions of the staple VAL-C
atmospheric model \citep{hudson-1981ApJS...45..635V}\footnote{The
VAL-C parameters are available within SolarSoft as the procedure
VAL\_C\_MODEL.PRO.}.  This is a ``semi-empirical model'' in which
interprets a set of observations in terms of the theory of radiative
transfer, but without any effort to have self-consistent physics.
Such a model can accurately represent the spectrum but may or may not
provide a good starting point for physical analysis.  Because the
optical depth of a spectral feature is the key parameter determining
its structure, one often sees the model parameters plotted against
continuum optical depth $\tau_{5000}$ evaluated at 5000\AA.  Just for
illustration, Figure~\ref{hudson-fig:ztausigma} shows the VAL-C
temperature separately as a function of height, column mass, and
optical depth.  Note that features prominent in one display may appear
to be negligible in another

The VAL-C model is an ``average quiet Sun'' model, and like all
static~1D models, it cannot describe the variability of the physical
parameters that theory and observation require (see other papers in
these proceedings, e.g., Carlsson's review).  Thus we should regard
the plasma parameters estimated here as order-of-magnitude estimates
only and note especially that the vertical scales, which depend in the
model on the inferred optical-depth scale, may be systematically
displaced.

The VAL-C model explicitly does not represent a chromosphere perturbed
by a flare.  \citet{hudson-1981ApJS...45..635V} and many other authors
give more appropriate models derived by similar techniques for flares
as well as other structures.  As the discussion of the $\gamma$-ray
signatures in Section~\ref{hudson-sec:gamma} suggests, though, a
powerful flare may be able distort the lower solar atmosphere
essentially beyond all recognition (especially in the impulsive
phase).  To estimate representative plasma parameters I have therefore
chosen just to start with the basic VAL-C model, and we simply assume
constant values of $B$ at 10~G and 1000~G.  The actual magnetic field
may vary through this region (the ``canopy'') but the details are
little-understood.  The $\gamma$-ray literature usually uses a
parametrization of the magnetic field strength $B \propto
P_{g}^{\alpha}$ (\cite{hudson-1983ApJ...264..648Z}), where $P_{g}$ is
the gas pressure.

\begin{figure}
\centering \includegraphics[width=0.95\linewidth]{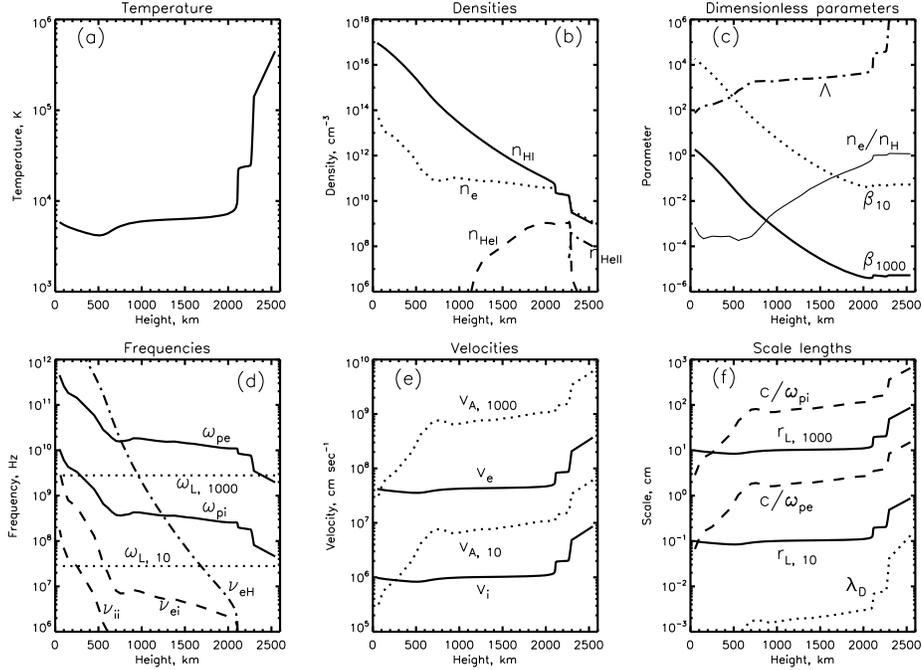}
\caption{Various plasma parameters in the VAL-C model.
We have assumed representative $B$ values of 10~G and 1000~G.
The different panels show the following, left to right and top to bottom:
(a) Temperature.
(b) Densities: solid, total hydrogen density; dotted, electron  density; dashed, He~I density; dash-dot, He~II density.
(c) Plasma beta: solid, for 1000~G; dotted, for 10~G; light solid, electron density as a fraction of total hydrogen density; dash-dot, the plasma parameter.
(d) Frequencies.  Solid, electron and ion plasma frequencies; dotted, electron gyrofrequencies for 10~and 1000~G; 
dashed, electron and ion collision frequencies; dash-dot, electron/neutral collision frequency.
(e) Velocities: Solid, electron and ion thermal velocities; dashed, Alfv{\' e}n speeds for 10~and 1000~G.
(f) Scale lengths: solid, electron Larmor radii for 10~G and 100~G; dotted, Debye length; dashed,
ion and electron inertial lengths.
}
\label{hudson-fig:chr_par}
\end{figure}

The most complicated behavior of the plasma parameters happens
preferentially near the top of the VAL-C model range (for example,
Figure~\ref{hudson-fig:chr_par} shows that the collision frequency
decreases below the plasma and Larmor frequencies) above the helium
ionization level (or even below this level for strong magnetic
fields).  Because VAL-C ignores time dependences and 3D structure, and
assumes $T_{e} = T_{i}$, we can expect that it has diminished fidelity
as one approaches the unstable transition region; thus one should be
especially careful not to take these approximations too literally.
The following notes correspond to each panel of the figure.  Most of
the plasma-physics formulae used in this Appendix are from
\citet{hudson-1984itpp.book.....C}.


\bigskip\noindent
{\bf Temperature:} The VAL-C model, like all of the semi-empirical models, sets T$_{e}$~=~T$_{i}$.
It therefore cannot support plasma processes dependent upon different ion and electron temperatures, or more complicated particle distribution functions \citep[e.g.][]{hudson-1994ApJ...427..446S}.

\bigskip\noindent {\bf Densities:} Total hydrogen density, electron
density, and densities of He\,I and He\,II.

\bigskip\noindent
{\bf Dimensionless parameters:} We approximate the plasma beta as $$\beta = {{{2(n_{H} + 2n_{e})kT}}\over{B^{2}/8\pi}}$$ with $n_{H}$ the hydrogen density, $n_{e}$ the electron density,
Figure~\ref{hudson-fig:chr_par}(c) gives the number of electrons in a Debye sphere as the  ``plasma parameter'' $\Lambda$.

\bigskip\noindent
{\bf Frequencies:} The plasma frequency, the electron and proton Larmor frequencies, and the
electron and ion and collision frequencies 
$$\nu_{ei} = 2.4 \times 10^{-6}  n_{e} ln\Lambda/T_{eV}^{3/2};\ \ \ 
\nu_{ii} = 0.05 \times  4\nu_{ei}; \ \ \ \nu_{eH} = (n_{H}/n_{e})\nu_{e}$$
with $n_{e}$ in cm$^{-3}$, $T_{eV}$ the temperature in eV,  using Z = 1.2 and  the Coulomb logarithm
ln$\Lambda$ =  23 - ln(n$_{e}^{0.5}$T$_{eV}^{-1.5}$)
\citep{hudson-1984itpp.book.....C,hudson-2001ApJ...558..859D}.

Note that the collision frequencies are small compared with the plasma and Larmor frequencies above about 1000~km in this model.
This means generally that plasma processes must have strong effects on the physical parameters of the
atmosphere in this region.

\bigskip\noindent
{\bf Velocities:} Electron and proton thermal velocities; Alfv{\'e}n speeds $v_{A}$ for 10~and 1000~G.

\bigskip\noindent
{\bf Scale lengths:} Electron Larmor radii for 10~and 1000~G, the ion inertial length $c/\omega_{pi}$, the electron inertial length $c/\omega_{pe}$, and the Debye length $\lambda_{D}$.
The inertial lengths determines the scale for the particle demagnetization necessary for magnetic reconnection.
For VAL-C parameters the ion inertial length increases to about 100~m in the transition region.




\end{document}